\documentclass[final,5p,times,twocolumn]{elsarticle}
\biboptions{numbers,sort&compress}

\usepackage{graphicx}
\usepackage{xcolor}
\usepackage{amsmath}
\usepackage{amssymb} 
\usepackage[section]{placeins} % This keeps floats inside their own section
\usepackage{upgreek} % Need this because particle names are labels therefore should NOT be italic
\usepackage[switch]{lineno}

\newcommand{\Klong}{\ensuremath{K_{\mathrm{L}}}}
\newcommand{\Kshort}{\ensuremath{K_{\mathrm{S}}}}
\newcommand{\Kshortlong}{\ensuremath{K_{\mathrm{S,L}}}}
\newcommand{\pion}{\ensuremath{\pi}}
\newcommand{\piplus}{\ensuremath{\pi^{+}}}
\newcommand{\piminus}{\ensuremath{\pi^{-}}}
\newcommand{\piplusminus}{\ensuremath{\pi^{\pm}}}
\newcommand{\pizero}{\ensuremath{\pi^{0}}}
\newcommand{\muon}{\ensuremath{\mu}}
\newcommand{\muminusplus}{\ensuremath{\mu^{\mp}}}
\newcommand{\neutrino}{\ensuremath{\nu}}

\journal{Physics of the Dark Universe}

\begin{document} 

\begin{frontmatter}

%\title{Measuring the level of CP~violation in space}
\title{Measuring gravitational effects on antimatter in space}
% \title{Measuring the gravitational behavior of antimatter in space}
% \title{CP~violation as a probe of antimatter gravitation}

\author[molise,lecce,milano]{G.M.~Piacentino}
\ead{nanni@fnal.gov}
\author[lnf,bu]{A.~Palladino}
\ead{palladin@bu.edu}
\author[lnf]{G.~Venanzoni}
\ead{graziano.venanzoni@lnf.infn.it}
\address[molise]{Universit\`a degli Studi del Molise, Campobasso, Italy}
\address[lecce]{INFN, Sezione di Lecce, Lecce, Italy}
\address[milano]{INAF, Sezione di Milano, Milano, Italy}
\address[lnf]{Laboratori Nazionali di Frascati dell'INFN, Frascati, Italy}
\address[bu]{Boston University, Boston, USA}
\date{March 27, 2015}
%\date{\today}

\begin{abstract}
%Repulsive gravitational effects between matter and antimatter 
%can lead to several interesting features, including an explanation for missing mass (dark matter), 
%the acceleration of the universe (dark energy), and the cosmic baryon asymmetry.
We propose an experimental test of the gravitational interaction with antimatter by measuring the branching 
fraction of the CP~violating decay $\Klong \to \piplus \piminus$ 
in space. We show that at the altitude of the International Space Station, 
gravitational effects may change the level of CP~violation such that 
a 5$\sigma$ discrimination may be obtained by collecting the $\Klong$ 
produced by the cosmic proton flux within a few years.
\end{abstract}


\end{frontmatter}

% \modulolinenumbers[5]
% \linenumbers

\section{Introduction}

The hypothesis that antimatter could have a different coupling to the gravitational 
field has fascinated physicists since the discovery of the first antiparticles. 
Various theoretical efforts have been put forth to demonstrate both the possibility of a 
different behavior or, on the contrary, the necessity of equal coupling. 
%old sentence before referee's request to change it:
% Several authors have shown that the possible gravitational repulsion 
% between matter and antimatter would offer a simple and elegant 
% explanation for a number of cosmological problems~\cite{chardin-1993,chardin_rax-1992}. 
%New sentence after referee's request to change it
Several authors have shown that the possible gravitational repulsion 
between matter and antimatter could offer at least a partial 
explanation for a number of cosmological problems, including those connected to dark matter 
and dark energy~\cite{chardin-1993,chardin_rax-1992,levy-chardin-2012,levy-chardin-2014,hajdukovic-2011-08-DM1,hajdukovic-2011-08-DM2,
hajdukovic-2012-01,hajdukovic-2012-05,hajdukovic-2014-04,hajdukovic-2016,villata-2013,blanchet-2007,blanchet-2008,blanchet-2009,bernard-2015}.

% \textcolor{red}{Add a new paragraph here. Is repulsive gravity allowed within the framework of GR? 
% Assumption in community that mass cannot be negative.
% Can we show that the stress--energy tensor is conserved(?) therefore eliminating the non-conservation
% of the source of gravitons, as indicated by the referee.
% Repulsive gravity is accepted in the case of wormholes or inflation (references?). 
% Kerr and Kerr-Newman metrics allow (require?) repulsive gravity.
% Refute ref's comment that describing DM and DE by repulsive gravity contradicts big bang nucleosythesis.}

At present, the state-of-the-art is not very different from the framework summarized in the 
article by Nieto et al.~in 1991~\cite{nieto_goldman-1991}. 
Limits on repulsive gravity have been calculated based on  
measurements~\cite{alves, fischler}.
%----Old before modifiying at referee's request----
% A relatively large number of experiments on the gravitational interaction of antimatter have been proposed and 
% even started, e.g. AEGIS~\cite{aegis}, ALPHA~\cite{alpha}, 
% ATRAP~\cite{atrap}, GBAR~\cite{GBar}, and a new proposal
% to measure a repulsive gravitational interaction involving the free fall of muonium atoms at PSI~\cite{kirsch,kaplan}.
% Most of them were related with direct measurement of the interaction of anti-protons 
% and of anti-nuclei and have to face technical difficulties due to the copious electromagnetic 
% noise and the difficulties of producing and confining anti-atoms. 
% Ultimately the only measurements related to the topic at hand came from
% CPLEAR in 1999~\cite{cplear-1999} and KLOE in 2000~\cite{mambriani_trentadue-2000} 
% where they looked for a modulation of CP~violation due to gravitational tidal 
% contributions from the moon, the Sun, and the galaxy.
%----New after modifications per referee's request----
A  relatively large number of experiments on the gravitational interaction of antimatter have been proposed and 
even started, e.g. AEGIS~\cite{aegis}, ALPHA~\cite{alpha}, 
ATRAP~\cite{atrap}, and GBAR~\cite{gBar}. 
%These experiments are in an advanced stage of R\&D and will be realized at CERN in the near future. 
In addition, the muonium experiment proposed at PSI~\cite{kirsch,kaplan} %is interesting because it 
is the first to involve a second generation fermion. Furthermore, aside from direct measurements 
in laboratories there are emerging astronomical tests as 
pointed out by~\cite{hajdukovic-2014-04,hajdukovic-2016} and supported by a
feasibility study~\cite{gai2014} on a trans-Neptunian Binary System. 
Our proposal of measuring CP~violation in a weaker gravitational field is complementary to both
laboratory experiments and astronomical observations; it is simply a direct test of 
the dependence of CP~violation on the gravitational field. Should such a dependence exist it would
be a strong indication of particles and antiparticles having opposite coupling with gravity.
Ultimately the only measurements related to the topic at hand came from
CPLEAR in 1999~\cite{cplear-1999} and proposed at KLOE in 2000~\cite{mambriani_trentadue-2000} 
where they looked for a modulation of CP~violation due to gravitational tidal 
contributions from the Moon, the Sun, and the galaxy.
% The CPLEAR collaboration concluded: 
% ``it has been suggested previously [...] that all the CP~violation observed in the 
% neutral-kaon mass matrix might be due to the interaction with an astrophysical source. 
% Our results, on the absence of modulations correlated, e.g., with the Earth-Sun distance 
% do not allow us to reject this hypothesis''~\cite{cplear-1999}.

The experiment proposed in this article is based on Good's argument~\cite{good-1961}. 
Good initially noted that the absence of CP~violation in neutral kaon decays would experimentally 
rule out any possible gravitational repulsion between matter and antimatter. 
%In fact before the discovery of CP~violation, Good had noticed that due to an absolute gravitational potential, 
%the energy difference between $\Kzero$ and $\KzeroBar$ due to antigravity ($\Delta E = 2 m_{\Kzero} \phi G$) would be reflected 
%in a difference of their time evolution, by instantly regenerating a $\Kshort$ from a $\Klong$. 
% In fact, the temporal 
% variation between the components of $\mathrm{K}_{0}$ and $\bar{\mathrm{K}}_{0}$
% is $e^{i 2m_{\mathrm{K}_{0}} \phi G t / h}$. 
The discovery of CP~violation has greatly limited the validity of Good's argument and various authors  
have considered variants of it both in favour and in opposition to the possible existence of a 
different gravitational coupling between matter and antimatter. 

This topic deserves experimental tests 
and current investigations are ongoing in various laboratories on Earth.
We instead propose to study a possible dependence of CP~violation 
on the gravitational interaction %and/or the violation of the weak equivalence principle 
in the $\Kshort$-$\Klong$ system in space. 
The magnitude of any difference between the CP~violation parameter, $\varepsilon$,
measured in orbit and that measured 
on Earth's surface would give important indication on the nature of the gravitational interaction between 
matter and antimatter as well as provide evidence for a quantum gravitational effect.
In this paper we outline a new approach to the problem capable of providing a 
5$\sigma$~measurement.

%=======================================================================================
\FloatBarrier
\section{Possible experimental setup}

%----old before modifications per referee----
% The mean gravitational field strength at the orbit of the 
% International Space Station (ISS) is about 
% 10\% less than on the Earth's surface. The orbit is largely circular so that the mean value of the local $g$
% is stable. During orbit the tidal effects of the gravitational field of the moon, sun, and galaxy 
% are due to distance variations of the order of $1.3 \times 10^7$~m.  
% The AMS-01 experiment on the ISS has measured the rate of incoming 
% protons~\cite{ams01_PLB427,ams01_PLB490} that when integrated on the permitted entrance to 
% the detector reaches as many as $2.2 \times 10^4$ protons per second. 
%---new after modifications per referee----
The mean gravitational field strength in a Low Earth Orbit (LEO) is about 
10\% less than on the Earth's surface. Following Chardin~\cite{chardin-1993,chardin_rax-1992} we
consider a dependence of $\varepsilon$ only on the local acceleration due to gravity, $g$, 
not on the gravitational potential. In circular orbits, such as a LEO, 
$(g_\mathrm{orbit} - g_\mathrm{surface})/g_\mathrm{surface}$ is stable and all external perturbations are at least an
order-of-magnitude less and will not be considered in this paper.
The rate of incoming protons in a LEO has been measured~\cite{ams01_PLB427,ams01_PLB490}, 
and when integrated on the permitted entrance to 
the detector can reach as many as $2.2 \times 10^4$ protons per second.
%---
The energy of the cosmic protons ranges from a few 
MeV to $\sim$200~GeV with the maximum flux
around 1~GeV and several smaller local maxima at 5, 13, and 31~GeV.

The incident proton spectrum is energetic enough for the 
production of $\Klong$ allowing for a measurement of 
\begin{equation}
R = \frac{\Gamma(\Klong \to \piplus\piminus)}{\Gamma(\Klong \to \piplus\piminus\pizero)}
\end{equation}
which is quadratic in $\varepsilon$.
If CP~violation depends linearly on the gravitational field~\cite{chardin-1993,chardin_rax-1992}
we expect a 10\% effect in $\varepsilon$ to translate to a 20\% effect on $R$.

\begin{figure}[t]
\begin{minipage}[t]{0.48\textwidth}
\centering
\includegraphics[width=.98\textwidth,trim=0 0 0 0,clip]{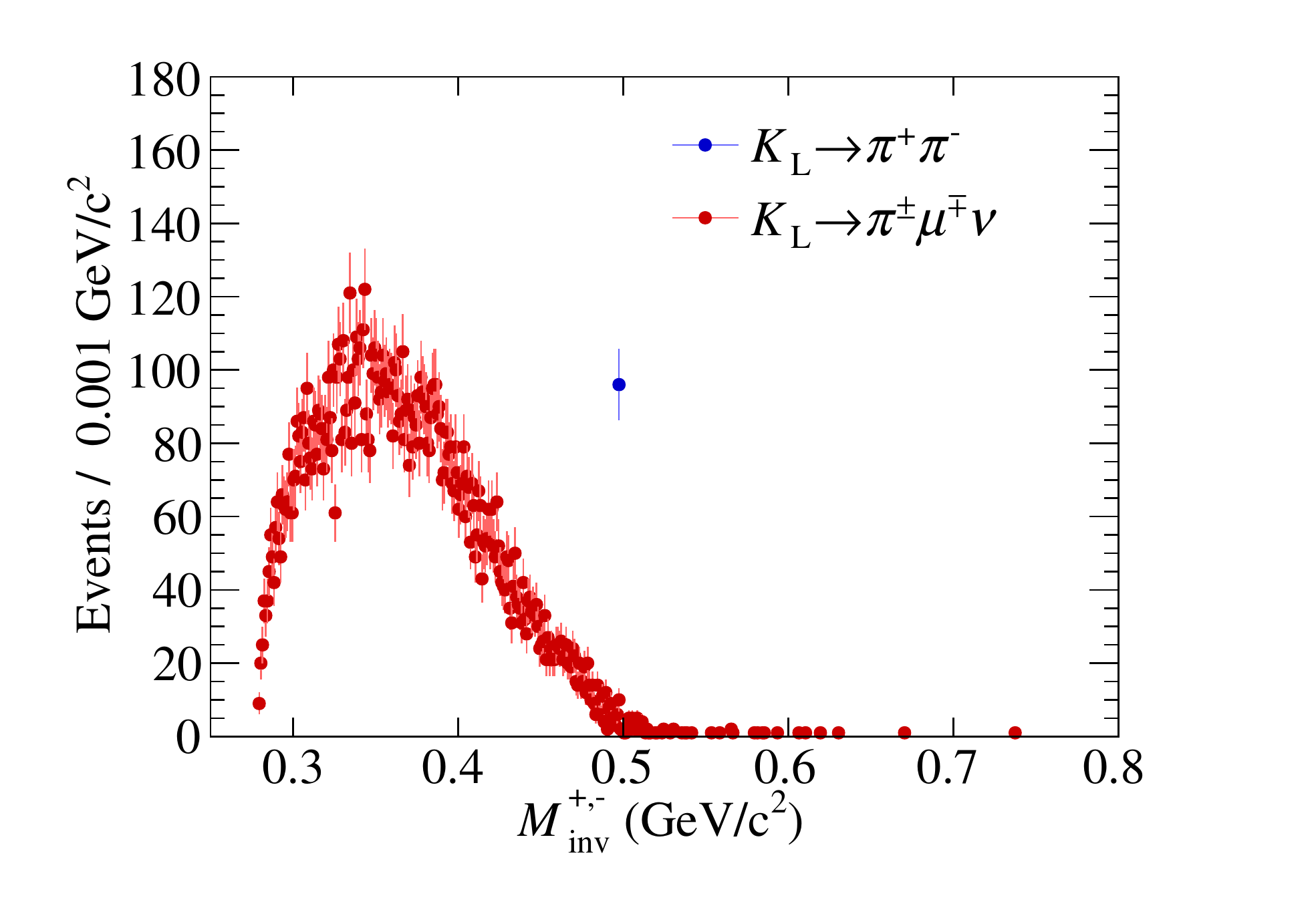}
\caption{\label{fig:KLdaughtersMinv} (Color online) Invariant mass calculated from the charged daughters of the 
$\Klong$ decays. For the 2-body decay we recover $m_{\Klong}$.
 A fit to this distribution and other kinematical observables can easily provide knowledge of 
 our background contributions with an uncertainty at the percent level, $\delta B/S < 0.02$.}
\end{minipage}
\end{figure}

To make a $5\sigma$ measurement of a 20\% effect on $R$ we would need to record at least
$12.5 \times 10^{5}$ $\Klong$~decays, with 
$N\left(\Kshort\text{ decays} \right) / N\left(\Klong\text{ decays} \right)< 5.7 \times 10^{-5}$,
while keeping the uncertainty in the background at less than 0.02 of our signal, $\delta B/S<2$\%.
The background contribution from $\Klong \to \piplusminus\muminusplus\neutrino$ can 
be easily obtained with an uncertainty at the percent level by fits to  
kinematical obervables such as the invariant mass calculated from the 
two charged tracks (Figure~\ref{fig:KLdaughtersMinv}).
The corresponding values necessary for a $3\sigma$ measurement are listed in Table~\ref{tab:sim_results}.
%----old before modifications per referee----
% As we will see in the following section this 
% yield can be achieved on the ISS within a few years of data taking.
%----new after modifications per referee----
As we will see in the following section this yield can be achieved on the 
proposed LEO within a few years of data taking.

%=======================================================================================
\FloatBarrier
\section{Simulation}

We performed a {\sc Geant4}~\cite{geant4_2003,geant4_2006} Monte Carlo simulation using the angular and 
energy spectrum of the incident cosmic protons as measured by AMS-01 spectrometer.
We simulated incident protons with $\theta_{\mathrm{max}}=45^\circ$ over a 
50~cm radius target surface corresponding to a $\pi/4$ solid-angle acceptance.

\begin{figure}[t]
\begin{minipage}[t]{0.48\textwidth}
\centering
\includegraphics[width=.98\textwidth,trim=0 0 0 0,clip]{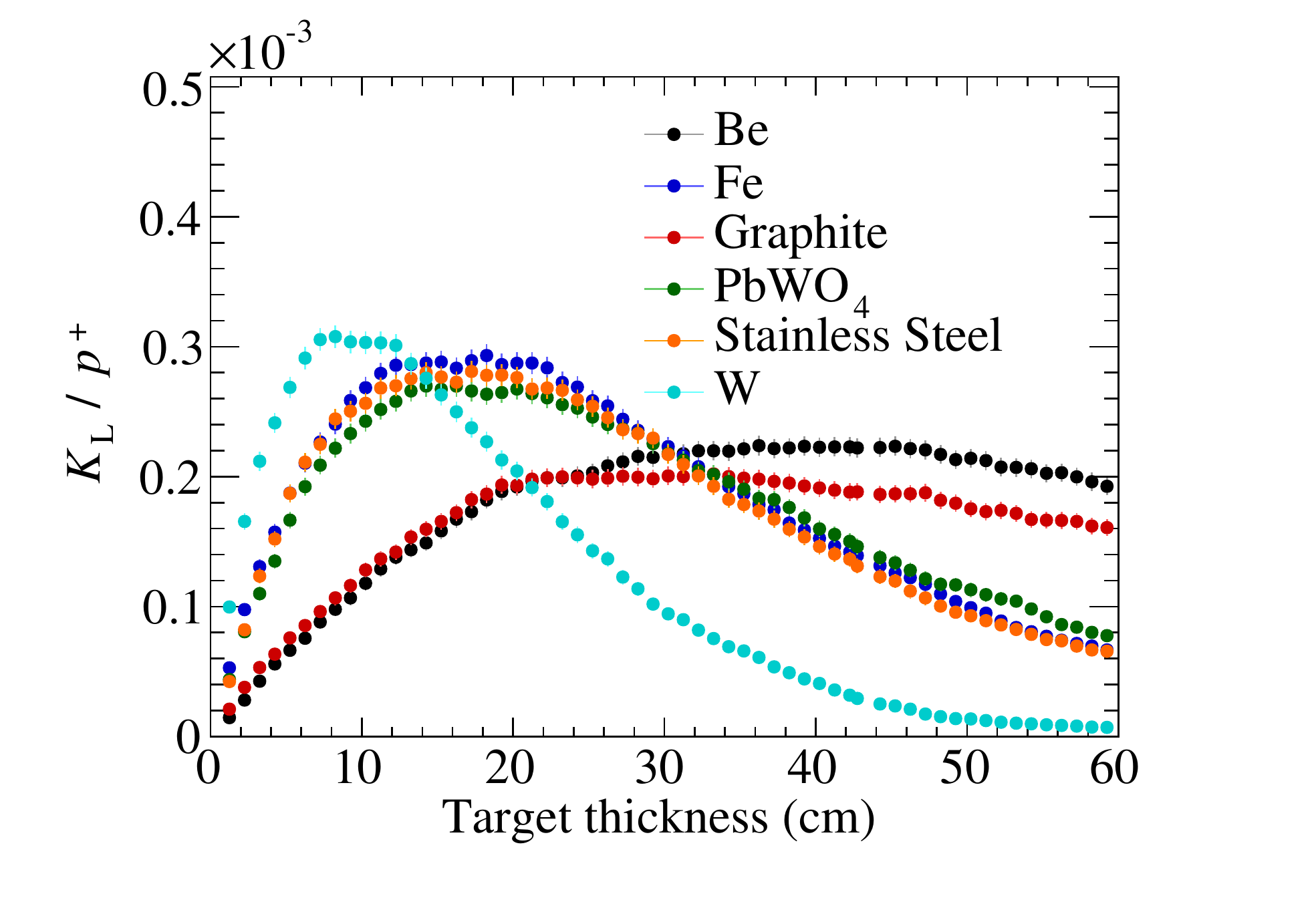}
\includegraphics[width=.98\textwidth,trim=0 0 0 0,clip]{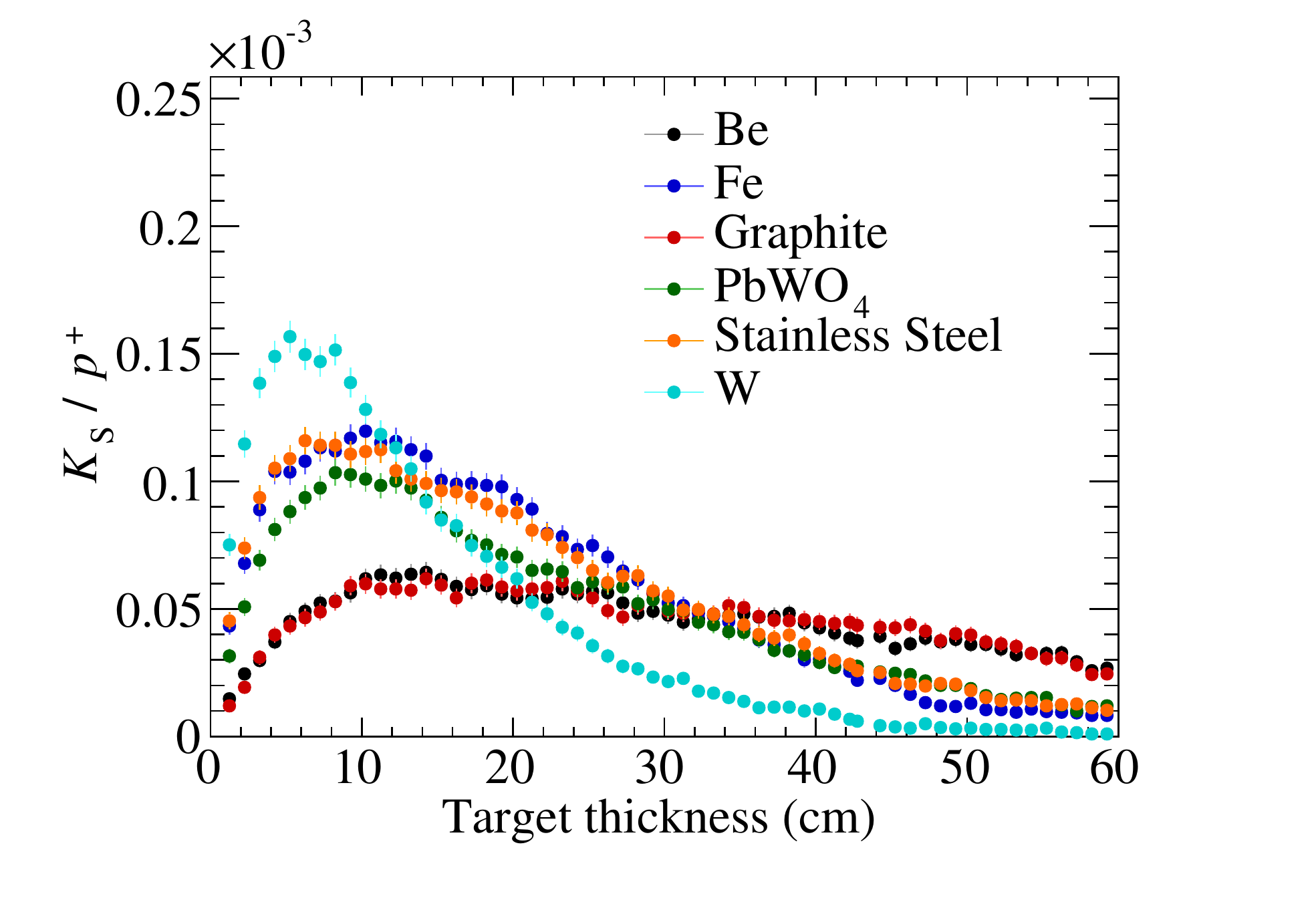}
\caption{\label{fig:BestThickness} (Color online) Number of $\Klong$ (\emph{top}) and 
$\Kshort$ (\emph{bottom}) which exit the downstream face of the target
versus target thickness for several materials. 
For our subsequent studies we used a 9~cm tungsten target which produced the most $\Klong$ 
with the shortest target thickness.}
\end{minipage}
\end{figure}

Figure \ref{fig:BestThickness} shows results from an optimization study of target material and depth. 
Even though more $\Klong$ are produced
for thicker targets, the probability that they exit the thicker targets is reduced due to regeneration and
nuclear interactions. 

\begin{figure}[t]
\begin{minipage}[t]{0.48\textwidth}
\centering
\includegraphics[width=.98\textwidth,trim=0 0 0 0,clip]{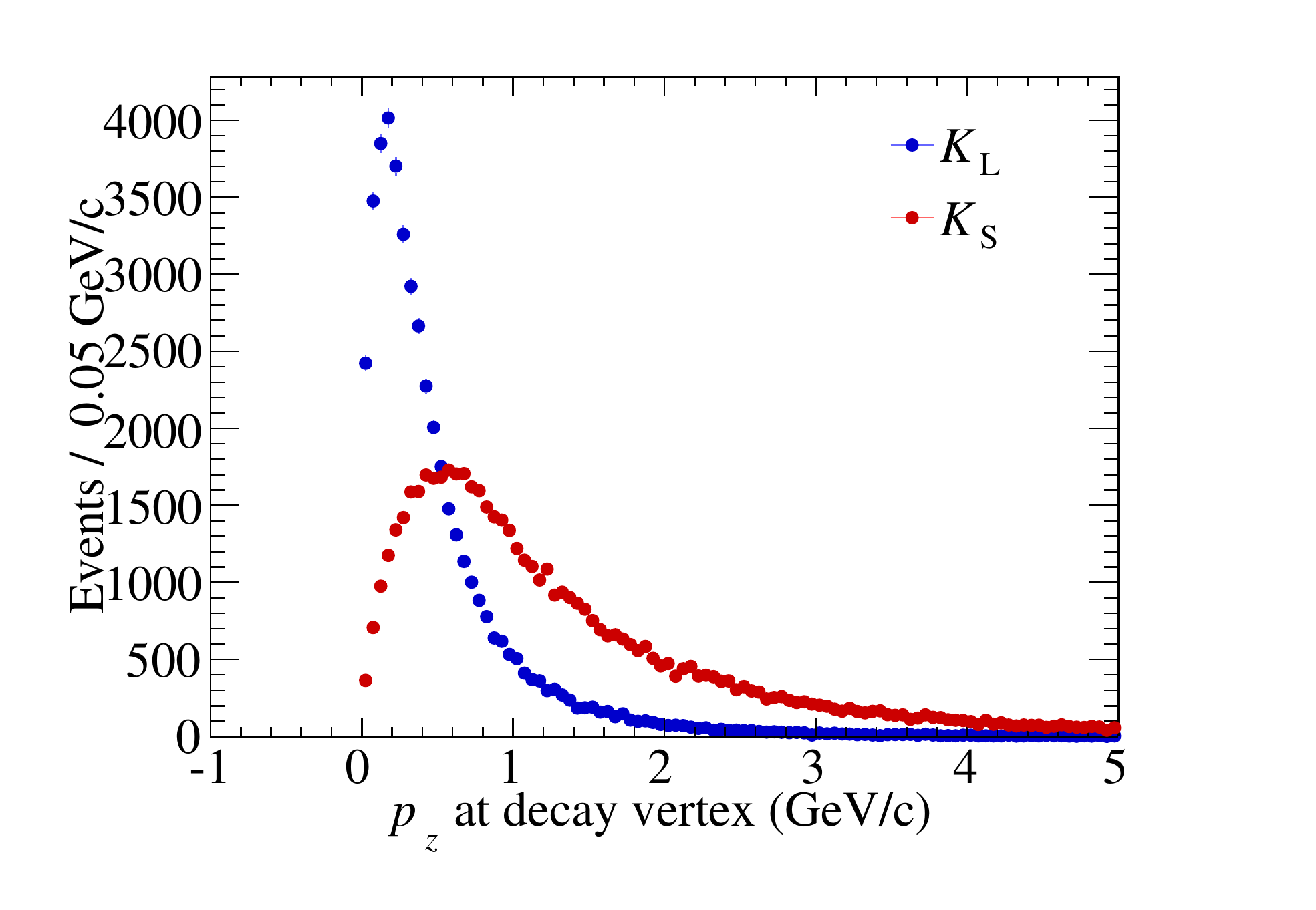}
\caption{\label{fig:Ek_exiting_target} (Color online) Axial momentum of the $\Kshortlong$ at their 
decay vertex, for the $\Kshortlong$ which decay in the tracking region 
downstream of the target, $z>0$.}
\end{minipage}
\end{figure}

\begin{figure}[t]
\begin{minipage}[t]{0.48\textwidth}
\centering
\includegraphics[width=.98\textwidth,trim=0 0 0 0,clip]{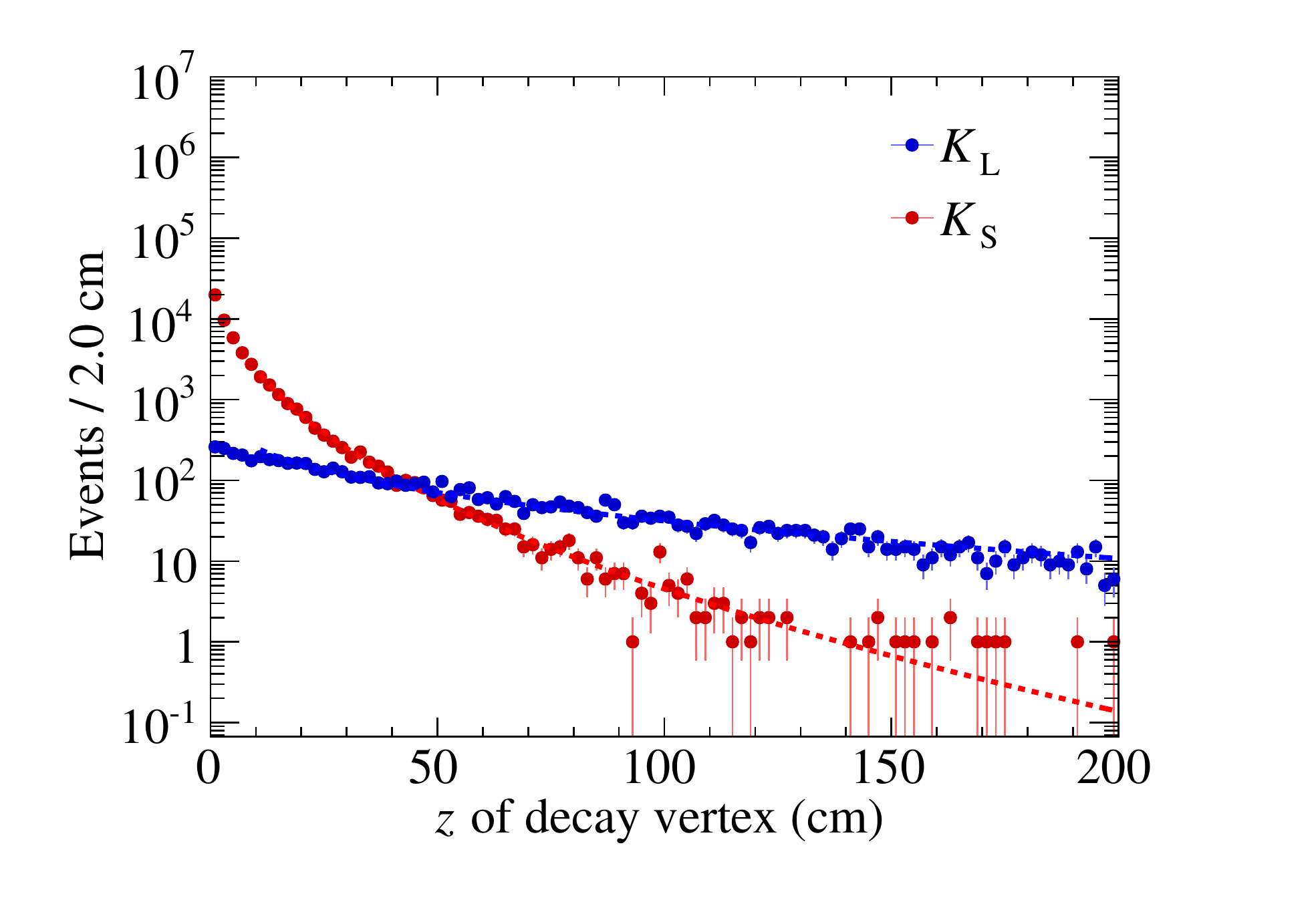}
\includegraphics[width=.98\textwidth,trim=0 0 0 0,clip]{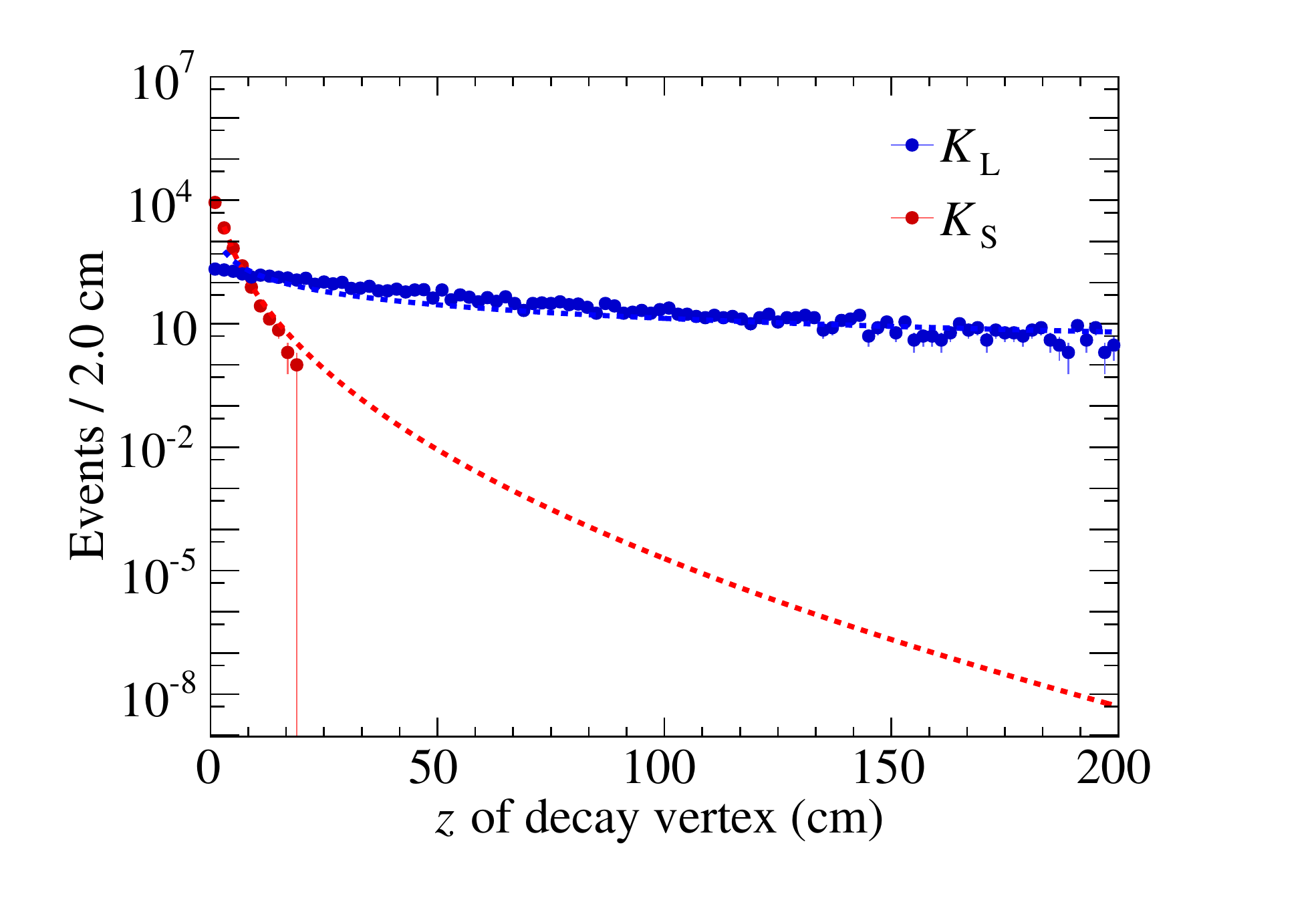}
\caption{\label{fig:z_of_decay} (Color online) Axial position of the $\Kshortlong$ decay vertex for 
$r<50$~cm (\emph{top}) and with the additional constraint $p_{z}<0.5$~GeV/c (\emph{bottom}).}
\end{minipage}
\end{figure}

%----old before modifications per referee----
% Using a tungsten target with a depth of 9~cm, and a radius of 50~cm, we studied the 
% $\Kshortlong$ that would decay inside the volume of a cylindrical tracking detector
% located downstream of the target. We found that the axial momentum ($p_{z}$) 
% distributions, Figure~\ref{fig:Ek_exiting_target}, for the 
% $\Klong$ and $\Kshort$ in the tracker differ such that we can suppress 
% the $\Kshort \to \piplus\piminus$ background with the cut $p_{z}<0.5$~GeV/c.
% Figure~\ref{fig:z_of_decay} shows the effect of the momentum cut.
% Considering a cylindrical tracking region with 1~m diameter, 1~m deep, 
% and offset 0.5~m downstream of the target,
% we would obtain the results given in Table~\ref{tab:sim_results}.
%----new after modifications per referee----
The detector geometry consists of a target composed of layers of tungsten and thin planes of active detectors 
with the tungsten having a cummulative depth of 9~cm and a radius of 50~cm.
Downstream of the target is a $\sim$50~cm deep charged-particle detector 
such as the Transition Radiation Detector (TRD) in AMS-02.
Next comes a cylindrical tracking volume followed at the downstream end by electromagnetic calorimeter.
The tracking region would be surrounded by a veto system to identify cosmic rays entering from the side.
The active layers sandwiched in the target would identify the parent incident proton and record the
time and position of the interaction point. Charged-particle backgrounds produced in the target would
be identified in the charged-particle detector between the target and the tracking region.
The neutral $\Kshortlong$ produced in the target, would travel into the cylindrical tracking region where they then decay.
Data analysis will select $\Klong \to \piplus\piminus$ and $\Klong \to \piplus\piminus\pizero$ events
with interaction vertices inside the tracking region which has a $>$50~cm displacement with respect to the
target thereby significantly reducing background contamination.
%inside the volume of a cylindrical tracking detector located downstream of the target. 
We found that the axial momentum ($p_{z}$) 
distributions, Figure~\ref{fig:Ek_exiting_target}, for the 
$\Klong$ and $\Kshort$ in the tracker differ such that we can suppress 
the $\Kshort \to \piplus\piminus$ background with the cut $p_{z}<0.5$~GeV/c. 
Figure~\ref{fig:z_of_decay} shows the effect of the momentum cut. 
Considering a cylindrical tracking region with 1~m diameter, 1~m deep, 
and offset 0.5~m downstream of the target, we would obtain the 
results given in Table~\ref{tab:sim_results}.
% We also considered and simulated other possible backgrounds that are presented 
% in Table 2. All of them can be subtracted thanks to the active target and to a  
% veto detector that tags events that have origin outside the target volume.

\newcommand\T{\rule{0pt}{3.1ex}}       % Top strut     %prevents superscripts from being hidden by hline above them.
\newcommand\B{\rule[-1.7ex]{0pt}{0pt}} % Bottom strut  %prevents  subscripts  from being hidden by hline below them.

%need 600 KL->2pi
%sqrt(600)=24.... 24/600 = 4%

% To make 5sigma meas at 20% we would require a 4% total error (conservatively 
% estimated at 2% stat error and 2% syst error),
% we conservatively estimate that we would need 
%  N KL needed for 5sig (2%+2%) is >12.5*10^5 while for 3sigma (4%stat+4%syst) is >3.0*10^5
%  NKs/NKL < 5.7*10^-5  at 5sigma and    NKs/NKL < 1.0*10^-4  at 3sigma
%  deltaB/S < 0.02 at 5sigma and < 0.04 at 3 sigma 

\begin{table}[h]
\centering
\caption{\label{tab:sim_results}Critical parameters necessary for $3\sigma$ and $5\sigma$ measurements of a 10\% change in the level 
of CP~violation (20\% change in $R$) along with the values obtained from our Monte Carlo simulation. The results take into account a 
basic geometrical event selection of $\Kshortlong$ decay vertices within a 1~m $\times$ 1~m cylindrical 
tracking volume 50~cm downstream of the target ($50<z<150$~cm, $r<50$~cm).
and axial momentum at the $\Kshortlong$ decay vertex of $p_{z}<0.5$~GeV/c.
These values assume a 100\% detection efficiency, 2\%~(4\%) statistical and 2\%~(4\%) systematic fractional
uncertainties for 5$\sigma$~(3$\sigma$).}
{\small
\begin{tabular}{c|c|c|c|c}
\hline \hline
{} & \multicolumn{2}{ c| }{ Requirement }       & \multicolumn{2}{ c }{ Simulation result }      \T\B \\
{} & 3$\sigma$      &  5$\sigma$ &  3$\sigma$      &  5$\sigma$  \\
\hline 
$             N\left( \Klong\text{ decays}     \right)$                                                            & $>3 \times 10^5$    & $>12.5\!\times10^5$   & 73 days  &  304 days  \T\B \\ %\hline  %\noalign{\vskip 1mm} 
$\frac{       N\left( \Kshort\text{ decays}     \right) }{ N\left( \Klong\text{ decays} \right) }$ & $<1 \times 10^{-4}$ & $<5.7\!\times10^{-5}$ & \multicolumn{2}{ c }{ $4.1 \times 10^{-5}$ }  \T\B \\ %\hline  %\noalign{\vskip 1mm} 
$\frac{\delta N\left( \Klong\to\pion\muon\neutrino \right) }{ N\left( \Klong\to\pion\pion  \right) }$ & $<4 \times 10^{-2}$ & $<2\!\times10^{-2}$   & \multicolumn{2}{ c }{ kinematical cuts      }  \T\B \\ \hline  %\noalign{\vskip 1mm} 
\hline
\end{tabular}
}
\end{table}

%=======================================================================================
\FloatBarrier
\section*{Conclusions}

We have proposed a possible test of the gravitational behavior of antimatter by measuring the  rate
of the CP~violating decay $\Klong \to \piplus\piminus$ in space.
We estimate that a 5$\sigma$ measurement on a 
possible change in the CP~violation parameter $\varepsilon$ could be obtained within a year, depending on the 
%final-state $\uppi^+$, $\uppi^-$, and $\uppi^0$ 
detection efficiency,
if one places a detector with 
a 9~cm thick tungsten target, a 1~m diameter by 1~m deep tracking region,
a magnetic field for charged-particle identification, time-of-flight counters, 
and electromagnetic calorimeters for energy measurements, on the International Space Station.
Any difference between the amount of CP~violation in orbit with respect
to the level CP~violation on the Earth's surface would be an indication 
of the nature of the gravitational interaction between matter and antimatter.
%and provide evidence for a quantum gravitational effect.
A positive result may offer an explanation for the cosmic baryon asymmetry 
%--old sentence before referee's request to change it:
%and may eliminate the need for dark matter and dark energy.
%--new sentence after changing it:
and may offer a contribution to the observed effects thought to come 
from dark matter and dark energy~\cite{chardin-1993,chardin_rax-1992,levy-chardin-2012,levy-chardin-2014,hajdukovic-2011-08-DM1,hajdukovic-2011-08-DM2,
hajdukovic-2012-01,hajdukovic-2012-05,hajdukovic-2014-04,hajdukovic-2016,villata-2013,blanchet-2007,blanchet-2008,blanchet-2009,bernard-2015}.
A negative result would also be of interest, confirming that the level of CP~violation
is independent of the absolute gravitational field.
Finally, we note that even if a target system is absent in existing satellite experiments 
such as AMS-02~\cite{ams-02} and Pamela~\cite{pamela}, 
it is possible that a number of $\Klong$ events have been recorded during 
data taking and are still available for analysis allowing for a first glance of the phenomenon.
%however a dedicated detector is necessary in order to reach a definitive result.

%=======================================================================================
\FloatBarrier
\section{Acknowledgements}
We thank G.~Chardin, D.S.~Hajdukovic, G.~Lamanna, E.~Recami, and G.F.~Bignami for useful discussions and suggestions.

%=======================================================================================
\FloatBarrier
\section{References}
\bibliographystyle{elsarticle-num} 
\bibliography{references}

\end{document}